\documentclass[aps,pra,twocolumn]{revtex4-1}
\pdfoutput=1
\usepackage{amsmath, amssymb}
\usepackage{bm}

\usepackage{color}

\usepackage{graphicx}
\usepackage{float}

\begin{document}
\title{Memory effects in the long-wave infrared avalanche ionization\\
of gases: A review of recent progress}
\author{E. M. Wright,$^1$ S. W. Koch,$^{1,2}$ M. Kolesik,$^1$ and J. V. Moloney$^{1,3}$}
\affiliation{$^1$College of Optical Sciences, University of Arizona, Tucson, Arizona 85721, USA}
\affiliation{$^2$Department of Physics and Material Sciences Center, Phillips-University, 35032 Marburg, Germany}
\affiliation{$^3$ Arizona Center for Mathematical Sciences, Department of Mathematics, University of Arizona, Tucson, Arizona 85721, USA}
\begin{abstract}
There are currently intense efforts being directed towards extending the range and energy of long distance nonlinear pulse propagation in the atmosphere by moving to longer infrared wavelengths, with the purpose of mitigating the effects of turbulence.  In addition, picosecond and longer pulse durations are being used to increase the pulse energy.  While both of these tacks promise improvements in applications, such as remote sensing and directed energy, they open up fundamental issues regarding the standard model used to calculate the nonlinear optical properties of dilute gases.  Amongst these issues is that for longer wavelengths and longer pulse durations, exponential growth of the laser-generated electron density, the so-called avalanche ionization, can limit the propagation range via nonlinear absorption and plasma defocusing.  It is therefore important for the continued development of the field to assess the theory and role of avalanche ionization in gases for longer wavelengths.  Here, after an overview of the standard model, we present a microscopically motivated approach for the analysis of avalanche ionization in gases that extends beyond the standard model and we contend is key for deepening our understanding of long distance propagation at long infrared wavelengths.  Our new approach involves the mean electron kinetic energy, the plasma temperature, and the free electron density as dynamic variables. The rate of avalanche ionization is shown to depend on the full time history of the pulsed excitation, as opposed to the standard model in which the rate is proportional to the instantaneous intensity.  Furthermore, the new approach has the added benefit that it is no more computationally intensive than the standard one.  The resulting memory effects and some of their measurable physical consequences are demonstrated for the example of long-wavelength infrared avalanche ionization and long distance high-intensity pulse propagation in air.   Our hope is that this report in progress will stimulate further discussion that will elucidate the physics and simulation of avalanche ionization at long infrared wavelengths and advance the field. 
\end{abstract}


\date{\today}
\maketitle
\section{Introduction}
Avalanche ionization was discovered by Townsend in the early twentieth century and is a process in which free electrons in a gas are accelerated by an electric field and subsequently free up more electrons via impact with neutral atoms \cite{Town}.  This process can cause the gas to become an electrical conductor and also produce exponential growth of the electron density leading to avalanche breakdown.  Nowadays, avalanche ionization is routinely produced in gases using pulsed lasers, with the laser field playing the dual role of producing free electrons via laser-induced ionization, and also accelerating the electrons to energies exceeding the atomic ionization potential so that the electron-atom collisions can lead to electron multiplication, see Refs. \cite{Raizer1,Raizer2} for a comprehensive overview of the field.  The resulting laser-induced plasma can have large effects on the subsequent laser propagation \cite{CouMys07,BerSkuNut07}, including absorption due to avalanche breakdown or limitation in the attainable focused spot size due to plasma-induced defocusing \cite{Blo74}.

In the present paper, we are not so interested in the case of avalanche breakdown with near complete ionization of all atoms.  Rather we consider the less extreme situation with a small ionization fraction that can accompany long distance nonlinear propagation, or optical filamentation, in atmospheric pressure gases such as air.  Optical filamentation has attracted a great deal of interest as it promises the possibility of near-diffraction free propagation of femtosecond to picosecond duration pulses over kilometers scales in the atmosphere.  In early studies of femtosecond filament propagation in gases, avalanche ionization did not play a significant role since the pulse durations were shorter than the doubling time for the electron multiplication process.  More recently, however, there has been increasing interest in nonlinear propagation for picosecond pulses and near infrared (NIR) to long-wave infrared (LWIR) wavelengths \cite{SkuBer07}-\cite{SchKolWri17}: Here avalanche ionization can become more significant since picosecond timescales exceed the doubling time for electron multiplication, and the quiver energy of the electron in the field, which scales as the wavelength squared, can lead to large freed electron kinetic energies in comparison to infrared pulses.  Both of these factors call for a re-examination of the standard model of avalanche ionization theory in the long-wave infrared regime that has been off great utility for femtosecond infrared pulses.  Moreover, it is important to have a reliable theory of avalanche ionization that is at the same time computationally simple enough that it can be used for detailed modeling of strong-field pulse propagation over extended time- and length scales and thus for the quantitative analysis of experiments.

In this Report in Progress, we develop a microscopically based description of avalanche ionization in gases that captures the dominant physics and is simple enough that it can be incorporated into long distance pulse propagation codes. The guiding principle behind our theory is the distinction between direct light-induced effects in gases, including electron excitation and ponderomotive acceleration, and many-body interaction effects among the electrons, atoms and ions that are independent of the simultaneous presence of a light field. In Ref. \cite{NatPhot} we demonstrated that incorporating this theory in our propagation simulations was essential in capturing the propagation of megafilaments at $10~\mu$m wavelength in air, the key point being that the standard model of avalanche ionization could not explain the experiments.  Our hope is that this report in progress will stimulate discussion that will elucidate the physics and simulation of avalanche ionization at long infrared wavelengths and advance the field. 

The remainder of this paper is organized as follows: Section \ref{standard_model} provides a brief overview of the standard model of avalanche ionization, and in Sec. \ref{BME} we recap previous results on thermalization of hot electrons in dilute gases according to the quantum Boltzmann equation (BME) as a basis for our model.  Section \ref{HEM} develops our hot electron model of avalanche ionization including a discussion of the limit in which our model reduces to the standard one.  Some illustrations of our model are given in Sec. \ref{Results} where we highlight the memory effects that arise for the case of LWIR pulse propagation, this serving to show that our new model can be incorporated into long distance propagation simulations. Finally, Sec. \ref{issues} discusses remaining issues, and summary and conclusions are given in Sec. \ref{Summary}
\section{Overview of the standard model}\label{standard_model}
Microscopically, the collision rate between electrons and neutral atoms is proportional to the product of their number densities, $N$ and $(N_0 - N)$ with $N_0$ the density of atoms, times the corresponding probabilities that the final states are available for in-scattering. In the dilute limit considered here, the final states are always unoccupied so that the rate of avalanche ionization is described by the density rate equation
\begin{equation}\label{Neq}
{dN\over dt} = \nu N(N_0-N)+{dN\over dt}\bigg |_{source} + {dN\over dt}\bigg |_{loss},
\end{equation}
where $\nu(t)$ is the ionization rate coefficient, and ${dN\over dt}\big |_{source}, {dN\over dt}\big |_{loss}$ are source and loss terms for the electron density, respectively. The source term describes laser-induced ionization and can be due to, for example, multiphoton or tunneling ionization, or excitation-induced dephasing (EID), and the loss term can include electron-ion recombination, electron attachment, or diffusion \cite{SunCheRud11,PolPasDri12,PapBotGor14}.  

In the standard model of avalanche ionization via inverse bremsstrahlung the ionization rate coefficient is proportional to the light intensity \cite{Raizer1}-\cite{BerSkuNut07}
\begin{equation}\label{nu_eq}
\nu(t) = {\sigma\over N_0 E_p} I(t)  ,
\end{equation}
where the cross-section for inverse bremsstrahlung is
\begin{equation}\label{sigma}
\sigma = {e^2\tau_e\over c m_e \epsilon_0} {1\over (1+(\omega\tau_e)^2 )} , 
\end{equation}
with $\omega$ the optical field frequency, $E_p$ the ionization energy, $m_e$ the electron mass, and $\tau_e$ the inverse electron scattering rate which can depend on both the electron density and temperature. 

The standard model can be obtained by a couple of different routes: The first proceeds from the observation that the rate of Joule heating of the electrons in the optical field is essentially that in a static field, yielding the rate of change of absorbed energy $W$ per electron \cite{YabBlo72}
\begin{equation}\label{dWdt}
{dW\over dt} = {e^2\tau_e\over m_e} {1\over (1+(\omega\tau_e)^2 )}E^2(t)  .
\end{equation}
Then assuming that all absorbed optical energy is converted to freed electrons via impact ionization, and identifying
\begin{equation}\label{Ndiff}
{dN\over dt} = {N\over E_P} {dW\over dt} = {\sigma I(t) \over E_p}N,
\end{equation}
yields the standard model in the low ionization limit $N(t)<<N_0$ \cite{Boyd}.  A second approach to obtaining the standard model is via the Fokker-Planck equation for the energy resolved electron distribution function \cite{Rubenchik}.  In this more sophisticated approach, Eq. (\ref{dWdt}) arises from the drift term of the Fokker-Planck equation, and the standard model readily follows from calculating the equation of motion for the mean electron density from the electron distribution function.

The standard model in Eq.~(\ref{Ndiff}) may be solved in the low ionization limit, and ignoring the loss term, if an ultrashort pulse generates an initial seed density $N_i$ via laser-induced ionization on a time scale before avalanche ionization initiates.  Then the density subsequently evolves according to \cite{Raizer1,Raizer2,YabBlo72}
\begin{equation}
N(t) = N_i \exp \left [ {\sigma\over E_p} \int_{ }^t dt' I(t') \right ] .
\end{equation}
This expression encapsulates the exponential growth of the electron density that characterizes avalanche ionization, and as a consequence avalanche ionization occurs only during the pulse, i.e. when $I(t) \neq 0$.   At first sight, this notion has some merits since surely the field must be present to accelerate the electrons, but it is neither microscopically founded nor generally correct. Fundamentally, impact ionization is a many-body interaction involving electrons, atoms, and ions, so that the rate of avalanche ionization should not depend directly on the light-field intensity but rather on the state of the matter system via its excitation energy and density, which are in turn driven by the field. As a consequence of this sequential coupling, avalanche ionization does not instantaneously follow the intensity, leading to memory or transient effects. As a related example exhibiting such memory effects, for a laser-produced high temperature plasma in air it has been shown both theoretically and experimentally that the plasma density can continue to increase well after the pulse has passed due to the high energy electrons in the tail of the Maxwell distribution that have enough energy for impact ionization of neutral atoms \cite{FilComRom09,RomComFil10,GaoPatSch17}.  The distinction in the present case is that the mean kinetic energy of the freed electrons already is enough to produce impact ionization both during and after the pulse.

\section{Microscopic approach}\label{BME}
Here we present a model of avalanche ionization that involves as dynamic variables the free electron density $N$, the mean electron kinetic energy expressed as a nonequilibrium temperature $T_{kin}$, and the plasma temperature $T_{pl}$ which is the temperature of the quasi-equilibrium state against which the electronic system relaxes due to the Coulombic electron-electron, electron-ion, and electron-neutral scattering. Our discussion starts from a microscopic approach and we show how the dominant microscopic physics can be incorporated into a hot electron model.

Ideally, a fundamental theory starts from a fully quantum mechanical level where the system is described by the Hamiltonian
\begin{equation}\label{Ham}
H = H_{single atom} + H_{atom-light} + H_{Coulomb} + ... 
\end{equation}
where the first term describes the energetics of an isolated atom, the second one the interaction of the individual atoms with the light field, the third one the Coulombic interaction of the charge carriers in the total system and so on. Assuming that the energies and eigenstates of the single-atom problem are known, one can express the light-matter interaction part as a combination of light-induced transitions between those states plus the ponderomotive contribution. The Coulomb interaction part then describes all the interactions among electrons and ions of the different atoms constituting the many-particle system.  In this paper, we are interested in strong field ionization of dilute atomic gases. Hence, one might think that the many-body Coulomb interactions could be negligible due to the relatively large inter-atomic separation. However, even though this may be a good approximation for the bound electrons of the individual atoms, it is not valid for the strong-field ionized electrons in spatially delocalized continuum states. Here, as argued below, the Coulomb scattering effects turn out to be significant due to the near absence of any screening effects \cite{PasMolKoc12}.

Since a fully microscopic approach is not viable in combination with the computational requirements for space-time field propagation over long distances, one typically reduces the complexity using systematic approximation schemes. An example, where the many-particle part of the avalanche ionization problem has been reduced to the level of quantum Boltzmann equations (BME) has been presented in Ref. \cite{PasMolKoc12}.    
Generically, these BME are written for the Wigner function $f_W^e({\bf r},{\bf k},t)$ which is a function of space and momentum coordinates as well as time. The BME includes drift and diffusion terms, the interaction with external fields, collision integrals as well as ionization and recombination terms
\begin{eqnarray}
& & \left [
{\partial\over \partial t} +{\hbar{\bf r}\over m_e}\nabla_{\bf r} - {1\over\hbar}(\nabla V)\cdot \nabla_{\bf k} 
\right ]f_W^e({\bf r},{\bf k},t) \nonumber \\
& =& {\partial f_W^e\over\partial t} \bigg |_{collision} + {\partial f_W^e\over\partial t} \bigg |_{ionization} 
+ {\partial f_W^e\over\partial t} \bigg |_{recombination}   .
\end{eqnarray}
The interaction with the electromagnetic field enters through the Lorentz force $-\nabla V=q_e({\bf E}+{\bf v}\times {\bf B})$, with $q_e$ and $m_e$ the electron charge and mass, and $\hbar{\bf k}=m_e{\bf v}$.  

In Ref. \cite{PasMolKoc12}, the BME was solved for an idealized gas of dilute hydrogen atoms excited by a femtosecond pulse and the subsequent time dynamics of the electronic system was monitored. The quantum calculations of the femtosecond laser-induced ionization yield highly anisotropic electron and ion angular (momentum) distributions on femtosecond timescales.  After the femtosecond pulse had passed a quantum Monte-Carlo analysis of the subsequent many-body dynamics revealed two distinct relaxation steps, first to a nearly isotropic and hot nonequilibrium configuration, and then to a quasi-equilibrium configuration. The collective isotropic plasma state is reached on a picosecond timescale well after the ultrashort ionizing pulse has passed.  

In the BME simulations, electron-ion as well as electron-neutral collisions were retained, but the calculations showed that the thermal relaxation was dominated by the electron-electron Coulomb scattering. At a superficial level, the notion that electron-electron scattering can thermalize the hot electrons on a timescale of an inverse plasma frequency in dilute gases appears to be counter intuitive since the corresponding collision integral scales as ${\partial f\over \partial t}\big |_{collision}\propto V^2f^2$, with $f$ a characteristic k-space occupation and $V$ representing the Coulomb potential. Since $f<<1$ in dilute gases, one might think that the electron-electron scattering will be too weak. However, one should keep in mind that the carrier-carrier scattering probability scales with the square of the interaction matrix element, i.e. the Coulomb interaction matrix element $V$. In the dilute gas system, the Coulomb interaction is basically unscreened with the consequence that the product $V^2 f^2$ is quite comparable to that in dense condensed matter systems where electron-electron scattering is acknowledged to be very important. Hence, the main message of these systematic dilute gas BME simulations for our current analysis is that the hot electrons of temperature $T_{kin}$ are thermalized and approach the plasma temperature $T_{pl}$ on the time scale $\tau= {1\over f_{pl}}$, with $f_{pl}(N(t))={1\over 2\pi}\sqrt{{N(t)e^2\over m_e\epsilon_0}}$ the density-dependent plasma frequency \cite{PasMolKoc12}. 

Other approaches for the light-matter interaction include numerical solutions of the Fokker-Planck equation for the energy-dependent electron distribution function \cite{Rubenchik} as well as solutions of the Maxwell-Schr\"odinger-plasma (MASP) model \cite{LorCheZao12}. However, even for sub-millimeter propagation lengths the MASP model is at the edge of current computational capabilities. 

\section{Hot electron model} \label{HEM}

Whereas the direct numerical solution of the BME was possible for the situation of an idealized dilute hydrogen gas at ultrashort time scales, this is computationally very demanding for more realistic systems and certainly not feasible at all for conditions where long distance propagation effects for picosecond time-scale pulses are of interest. Hence, we need a microscopically based but computationally less demanding approximation scheme. For this purpose, we  have to further simplify the fully microscopic theory in order to deal with macroscopic variables rather than the Wigner function for the electronic system. 

Here, we develop a model where we describe avalanche ionization in terms of the electron density, the electron kinetic temperature, and the plasma temperature.  The electron density and electron kinetic temperature reflect the state of the electronic system whilst the plasma temperature is the temperature of the quasi-equilibrium state towards which the electronic system relaxes due to the Coulombic electron-electron, electron-ion, and electron-neutral scattering.  We also discuss the limit in which our model reduces to the standard model for avalanche ionization.

\subsection{Electron and plasma temperatures}
During avalanche ionization, newly liberated free electrons can acquire very large kinetic energies due to the laser pulse, thereby creating hot electrons \cite{GorPlaSte09}, whilst many-body Coulomb interactions work to drive the system towards a collective plasma state.  Here, we identify the mean electron kinetic energy of the hot electrons with the non-equilibrium temperature $T_{kin}$, along with the temperature $T_{pl}<T_{kin}$ of the collective plasma state.  Then, allowing for the fact that the electron kinetic temperature relaxes to the plasma temperature on the time scale $\tau= {1\over f_{pl}}$, the equation for the electron kinetic temperature becomes
\begin{equation}\label{Te}
{dT_{kin}\over dt} = - { (T_{kin}-T_{pl})\over\tau}  + {dT_{kin}\over dt}\bigg |_{aval}  + {dT_{kin}\over dt}\bigg |_{coll},
\end{equation}
where ${dT_{kin}\over dt}\big |_{aval}$ and ${dT_{kin}\over dt}\big |_{coll}$ are the temperature loss rates due to avalanche ionization, and electron-ion and electron-neutral collisions, respectively.  The corresponding equation for the plasma temperature is
\begin{equation}\label{Tpl}
{dT_{pl}\over dt} = -{ (T_{pl}-T_{kin})\over\tau}  + {dT_{pl}\over dt}\bigg |_{cool} ,
\end{equation}
where ${dT_{pl}\over dt}\big |_{cool}$ is the loss rate due to impact ionization cooling \cite{FilComRom09}. This equation expresses the fact that the many-body Coulomb interactions will establish a quasi-equilibrium state of the ionized electrons that can be described as a plasma with given density and temperature.  

For the time being, we neglect the collision terms and the impact ionization cooling term in order to focus on the effects related to avalanche ionization.  Then subtracting Eqs. (\ref{Te}) and (\ref{Tpl}), we obtain the loss rate of the temperature difference
\begin{equation}\label{diff}
{d(T_{kin}-T_{pl})\over dt}\bigg |_{loss} = -{ 2(T_{kin}-T_{pl})\over \tau} + {dT_{kin}\over dt}\bigg |_{aval} .
\end{equation}
Physically $k_B(T_{kin}-T_{pl})$ is the excess energy per electron that is available for energy-dependent processes such as avalanche ionization.

\subsection{Density equation}
Starting from Eq. (\ref{Neq}) for the electron density and neglecting the source and loss terms, we can use a linear approximation to the ionization rate coefficient
\begin{equation}
\nu(T_{kin},T_{pl},N) \approx \nu(T_{pl},T_{pl},N)  + \nu' (T_{kin}-T_{pl})   .
\end{equation}
The term $\nu(T_{pl}) \equiv \nu(T_{pl},T_{pl},N)$ is the ionization rate coefficient that underlies impact ionization cooling of the electron plasma as previously described in Refs. \cite{FilComRom09,RomComFil10}, whereas the second term proportional to $\nu' = {\partial \nu(T_{kin},T_{pl},N)\over \partial T_{kin}}$ describes avalanche ionization.  Here, we isolate the avalanche ionization contributions so the density rate equation becomes
\begin{equation}\label{Ndot}
{dN\over dt} = \nu' (T_{kin}-T_{pl}) N \left (N_0-N \right ) .
\end{equation}
This equation is solved assuming there is a background density $N_b$ of free electrons even in the absence of the laser pulse.  Here we set $N_b=10^{13}$  m$^{-3}$ although the numerical results presented below were robust against this value if it was not chosen too large.

\subsection{Temperature loss rate}
To determine the parameter $\nu'$ that controls the temperature loss rate due to avalanche ionization, we start from the observation that energy conservation demands that
\begin{equation}
{d\over dt} \left ( {3\over 2} N k_B(T_{kin}-T_{pl}) \right ) + E_p {dN\over dt} = 0,  
\end{equation}
where $E_p$ is the ionization energy, $T_{pl}$ the density-dependent plasma temperature, and we have made use of the fact that $k_B(T_{kin}-T_{pl})$ is the excess energy per electron available for avalanche ionization.  Using this conservation law leads to the alternate expression for the loss rate in Eq. (\ref{diff})
\begin{widetext}
\begin{equation}\label{Tl1}
{d(T_{kin}-T_{pl})\over dt}\bigg |_{loss} =  -\left ((T_{kin}-T_{pl})  + {2E_p\over 3k_B} \right ) {1\over N} {dN\over dt}
 =  -\left ((T_{kin}-T_{pl})  + {2E_p\over 3k_B} \right ) \nu' (T_{kin}-T_{pl})\left (N_0 -N \right ) .
 \end{equation}
 \end{widetext}
In the low excitation limit where $k_B(T_{kin}-T_{pl})<<E_p$, the avalanche ionization term ${dT_{kin}\over dt}\big |_{aval}$ should be negligible, in which case comparing Eqs. (\ref{Tl1}) and (\ref{diff}) yields
$\nu' = {3k_B\over E_p} {1\over \tau} {1\over (N_0-N)}$.  Then substituting this expression back into Eq. (\ref{Tl1}) and comparing to Eq. (\ref{diff}) without imposing the low excitation limit yields the electron temperature loss rate due to avalanche ionization
\begin{equation}
 {dT_{kin}\over dt}\bigg |_{aval} = - {3k_B(T_{kin}-T_{pl})^2 \over \tau E_p}  .
 \end{equation}

\subsection{Equations for avalanche ionization}
Combining the above results, including sources for laser-induced heating and ionization, and reintroducing the collision, loss, and impact ionization cooling terms previously neglected, the three equations for transient avalanche ionization become
\begin{widetext}
\begin{eqnarray}
{dT_{kin}\over dt} &=& {2e^2\tau_e I(t)\over 3k_B c\epsilon_0 m_e(1+(\omega\tau_e)^2)} - { (T_{kin}-T_{pl})\over\tau} 
- {3k_B(T_{kin}-T_{pl})^2 \over \tau E_p}  +{dT_{kin}\over dt}\bigg |_{coll} , \label{A} \\
{dT_{pl}\over dt} &=& -{ (T_{pl}-T_{kin})\over\tau} -\left (T_{pl}  + {2E_p\over 3k_B} \right )\nu(T_{pl})(N_0-N) , \label{B} \\
{dN\over dt} &=&   {N\over \tau} \cdot {3k_B(T_{kin}-T_{pl}) \over E_p} + \nu(T_{pl})N(N_0-N)  + {dN\over dt}\bigg |_{source} + {dN\over dt}\bigg |_{loss} \label{C} .
\end{eqnarray}
\end{widetext}
In the equilibrium case, $T_{kin}=T_{pl}$, and in the absence of the source and loss terms for the density, Eqs. (\ref{B}) and (\ref{C}) are identical in form to those derived by Romanov {\it et al.} \cite{RomComFil10} to describe impact ionization cooling.  The key development here with respect to Ref. \cite{RomComFil10} is the extension to include the hot electrons via their temperature Eq. (\ref{A}). In addition, the first term in Eq. (\ref{A}) describes heating of the free electrons by the laser pulse, with $\nu_e(N,T_{kin})={1\over\tau_e}$ the electron scattering rate, $k_B$ is Boltzmann's constant, and $\omega$ the center frequency of the applied field.  For the remainder of the discussion we shall again neglect the collision, loss, and impact ionization cooling terms and concentrate on the dominant avalanche ionization contributions.

\subsection{Reduction to standard model}

Our model approaches the standard model of avalanche ionization via inverse bremsstrahlung in the limit of high pressures compared to atmospheric, so that the densities and associated plasma frequencies ${1\over\tau}$ and scattering rates $\nu_e$ can be much larger, meaning that lower temperatures will be attained.  Then the kinetic temperature Eq. (\ref{A}) is dominated by the first two terms, and for long enough pulses, $t_p>\tau,1/\nu_e$, this may be solved in steady-state for $(T_{pl}-T_{kin})$ as a function of $I(t)$.   Substituting this solution into the first term of Eq. (\ref{C}) then yields 
\begin{equation}\label{standard}
{dN\over dt} = {\sigma\over E_p} I N ,
\end{equation}
where the cross-section for inverse bremsstrahlung $\sigma$ is the same as given in Eq. (\ref{sigma}).  That the standard model appears at high pressures is not surprising as the atomic system approaches densities characteristic of liquids and solids for which the standard model was originally developed \cite{Blo74}.

\subsection{Approach to equilibrium}
For low to medium high pressures, we see from Eq. (\ref{C}) that the rate of avalanche ionization is proportional to the temperature difference $(T_{kin}-T_{pl})$. Therefore, it is interesting to establish the time scale for the equilibration of the kinetic and plasma temperatures since that will determine the time scale $T$ over which avalanche ionization persists. 

In the absence of the field, the equation for the dimensionless temperature difference $\Delta=k_B(T_{kin}-T_{pl})/E_p$ can be written as
\begin{equation}\label{Delta}
{d\Delta\over d\xi} = -2\Delta - 3\Delta^2  , \quad \xi(t) = \int_{0}^t {dt'\over\tau(t')}  ,
\end{equation} 
with $\xi(t)$ the dimensionless time variable.  This equation is universal in the sense that it is the same for all gases, the material details being subsumed into the dimensionless quantities.  For the physically relevant case here with $\Delta(0)>>1$, there is an abundance of excess energy to contribute to avalanche ionization such that the nonlinear term in Eq. (\ref{Delta}) dominates, leading to the solution $\Delta(\xi)=\Delta(0)/(1+3\Delta(0)\xi)\approx {1\over 3\xi}$.  The timescale $T$ for equilibration may then be assessed from
\begin{equation}\label{xi}
\xi(T) = \int_{0}^T {dt'\over \tau(t')} = \int_{0}^T dt' f_{pl}(N(t')) \simeq 1 ,
\end{equation}
meaning that low electron densities and associated plasma frequencies will lead to long equilibration times.  As the coarsest approximation, this yields $T\propto\tau = {1\over f_{pl}}$, with $f_{pl}$ the maximum plasma frequency attained during the pulse.  However, this underestimates $T$ and we refer to the full Eq. (\ref{xi}) in the following discussion.

We also note that in terms of the same dimensionless parameters, the solution of Eq. (\ref{C}) with no source but an initial density $N(0)<<N_0$ may be written as
\begin{equation}
N(t) = N(0) \exp\left (\int_0^{\xi(t)} d\xi' \Delta (\xi') \right ) , 
\end{equation}
which reveals the exponential growth of the avalanche ionization when $\Delta=k_B(T_{kin}-T_{pl})/E_p>0$.  This generalizes previous results without memory effects \cite{Blo74}.

\section{Results and discussion}\label{Results}

In this section, we present simulations to elucidate some of the memory effects that can occur.  After a brief discussion of the simulation approach and parameter values in Sec. \ref{IVA}, Sec. \ref{IVB} describes the basic memory effects by considering a point model without propagation, and this serves as a basis for the subsequent discussion.  Specifically, in Sec. \ref{IVC} we discuss one-dimensional pulse propagation, as appropriate to a gas filled hollow fiber geometry, and in Sec. \ref{IVD} we consider a three-dimensional example.  The simulations presented are intended to serve the dual purpose of elucidating the basic memory effects and also demonstrating that the resulting material equations are sufficiently simple that they can be incorporated into long distance pulse propagation codes.

\subsection{Simulations}\label{IVA}

As an example of current interest, we consider the case of air at atmospheric pressure with $E_p\simeq 12$ eV, and the expressions for the density and temperature dependent electron scattering rates $\nu_e(N,T_{kin})$ given in Ref. \cite{SunCheRud11}.  The first term in the kinetic temperature Eq. (\ref{A}) shows that, all other factors being equal, the rate of heating varies as $\omega^{-2}$, this being related physically to the electron jitter or quiver energy. Since the rate of avalanche ionization is related to the difference between the kinetic and plasma temperatures, it is clear that the ionization and associated memory effects will be more pronounced at longer wavelengths.  For this reason we consider an incident LWIR pulse of center wavelength $10~\mu$m and peak input intensity not exceeding $I_p=1.2\times 10^{12}$ W/cm$^2$ in the simulations. This peak intensity is well below the threshold for tunneling ionization, and the dominant source of free electron generation in this regime, and at atmospheric pressure, is excitation-induced dephasing (EID) due to many-body Coulomb effects which enhance the low-intensity electron densities \cite{SchHadMol15}. We have previously developed a consistent fit to the microscopic source term ${dN\over dt}\big |_{source}$ in terms of the applied field to capture the EID effect \cite{SchKolWri17}.  More specifically, this accurate fit is provided by the equation
\begin{equation}
{dN\over dt}\Bigg |_{source} = C_{MBI} E^4(t) \sqrt{ {E^2(t) + s\over E^2(t)} }   ,
\end{equation}
where for the simulations presented here, we used $s=4.6\times 10^{18} ~{V^2\over m^2}$ and $C_{MBI} = 2.84\times 10^{-7}~{1\over m^3s}{m^4\over V^4}$.  This source term is used in the density equation (\ref{C}), and for these simulations we neglected the effects of collisions, loss, and impact ionization to isolate the dominant impact ionization contributions. 

\subsection{Point model simulations}\label{IVB}

For the point model simulations without propagation effects, we solve Eqs. (\ref{A})-(\ref{C}) with a peak intensity $I_p=1.2\times 10^{12}$ W/cm$^2$.  The Unidirectional Pulse Propagation Equation (UPPE) package was used to perform the one- and three-dimensional simulations over a medium length $L=1$ m.  This approach allows one to include any material response, and in addition to the linear optical properties of air we incorporated the nonlinear refractive-index change
\begin{equation}
\Delta n = n_2 I - N{e^2\mu_0\over 2m} {\lambda^2\over (2\pi)^2}  ,
\end{equation}
with electronic Kerr coefficient $n_2=5\times 10^{-23}$ W/m$^2$, the second term being the plasma contribution.   Furthermore, assuming the mean velocity of newly ionized electrons is zero we also incorporated the following electron current density model in the UPPE
\begin{equation}
{dJ\over dt} = N{e^2E\over m}  ,
\end{equation}
which accounts for the effect of plasma absorption \cite{SchKolWri17}.

\begin{figure}[t!]
\includegraphics[width=9cm]{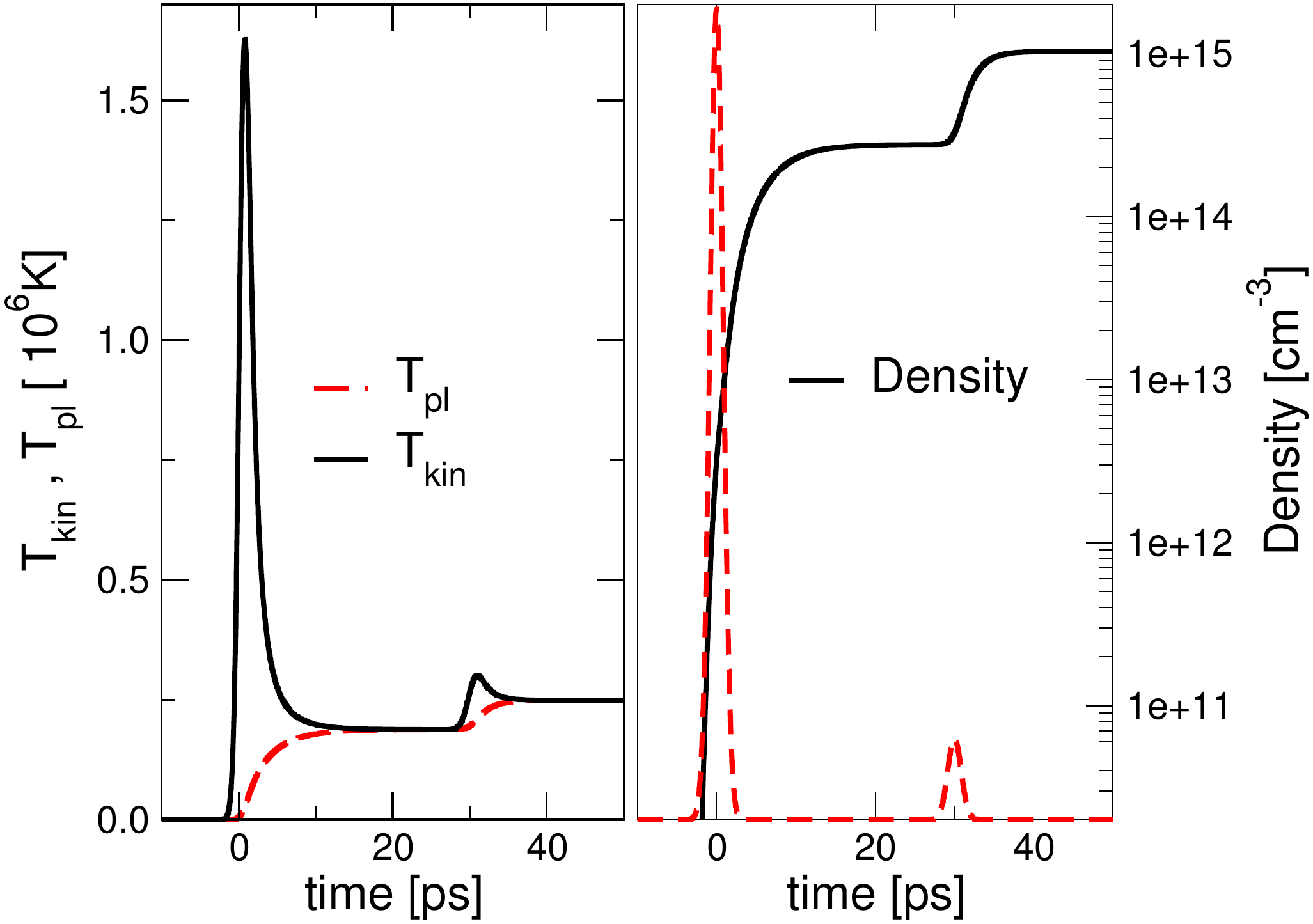}
\caption{(a) $T_{kin}$ (solid line) and $T_{pl}$ (dash line), and (b) the corresponding electron density (solid line) versus time, the dash line showing the intensity profile of the incident double-pulse.}\label{figure1}
\end{figure}

As an illustrative example, we show in Fig. \ref{figure1} the simulation results for two $2$ ps pulses (FWHM) separated by $30$ ps, the first (pump) pulse being centered on $t=0$, and the second (probe) pulse having one tenth of the peak intensity.  Figure \ref{figure1}(a) depicts $T_{kin}$ (solid line) and $T_{pl}$ (dash line) as functions of time, and we see that the temperatures approach equilibrium after the first pulse at about $T=10$ ps, in reasonable agreement with $T\approx 10.8$ ps from Eq. (\ref{xi}).  When the second pulse centered at $t=30$ ps arrives, a second boost in the two temperatures is clearly seen.  Figure \ref{figure1}(b) shows the corresponding electron density (solid line) versus time, the dash line representing the intensity profile of the incident double-pulse.  Here, the memory effect is clearly evident as the electron density continues to increase well after the first pulse has passed which is a key feature captured by our hot electron model.  Furthermore, when the second pulse arrives at $t=30$ ps, a second boost to the electron density arises. We see that due to the pre-existing or seed density, the timescale of the memory features is clearly shorter for the second pulse for which equilibrium is re-established after about $T=5$ ps as seen from Fig. \ref{figure1}(b), versus $T=4.2$ ps found from Eq. (\ref{xi}).  

Whereas the basic memory effect is that a strong initial pulse can create a seed density for the second (probe) pulse, we have the additional feature that the seed density depends on the delay time between the pulses.  This is distinct from the standard model in which the generation of electrons ceases between the pulses where the pulse intensity is zero.  In Sec. \ref{IVC}, we present simulations to demonstrate that the variation of seed density with delay can be measured by monitoring the probe transmission.

\begin{figure}[t!]
\includegraphics[width=8cm]{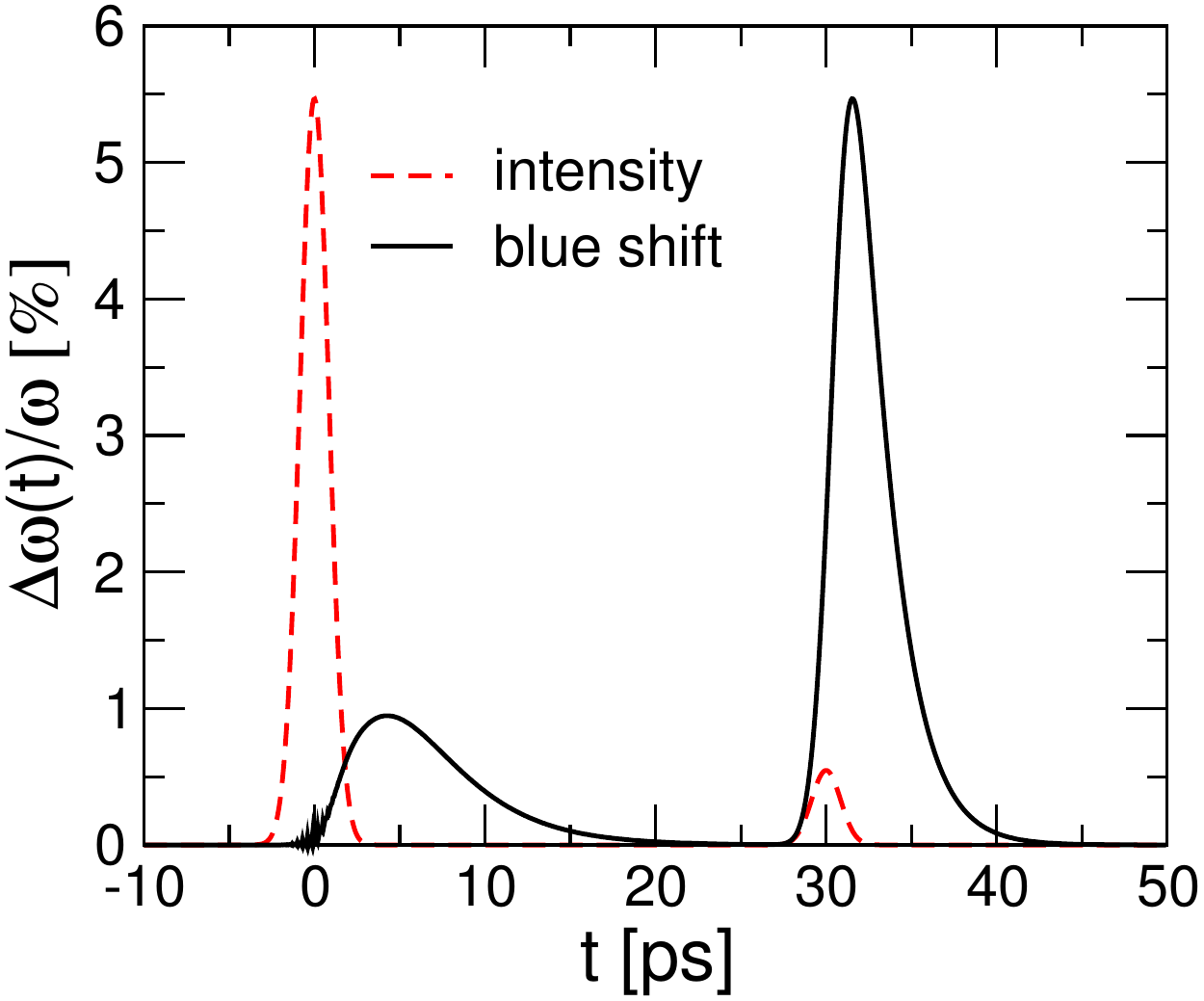}
\caption{Plasma blue-shift $\Delta\omega(t)$ versus time for the same case as Fig. \ref{figure1}, the dash line showing the intensity profile of the incident double-pulse.}\label{figure2}
\end{figure}

As another consequence of the memory effects, we discuss the feature of plasma blue-shifting.  The phase-shift of a plane-wave field after propagating through a thin gas cell of length $L$ may be approximated as $\phi(t)={\omega L\over c}\left (1-{N(t) \over N_c} \right ) -\omega t$, with $N_c={\epsilon_0m_e\omega^2\over e^2}$ the critical plasma density \cite{CouMys07}.  (Here we have neglected the Kerr nonlinear term to isolate the plasma contribution.)  The instantaneous frequency shift is then $\omega_{eff}(t)=-{\partial\phi\over\partial t}$, and the frequency shift is defined as \cite{Yab74,Yab88}
\begin{equation}\label{Del_om}
\Delta\omega(t)=\omega_{eff}(t)-\omega={\omega L\over c}{1\over N_c}{dN\over dt}  .
\end{equation}
For the same double-pulse as in Fig. \ref{figure1}, we have evaluated the scaled frequency shift ${\Delta\omega(t)\over\omega}$ as a percentage versus time using Eq. (\ref{C}) for ${dN\over dt}$. The results are shown in Fig. \ref{figure2} for $L=1$ m, the dash line representing the double-pulse intensity profile.  As expected, a plasma blue-shift occurs for all times \cite{Yab74,Yab88}. However, the peak of the plasma blue-shift associated with the first pulse centered around $t=0$ occurs at $t\approx 4$ ps, i.e., after the pump pulse has passed.  This is a consequence of the memory effects delaying the nonlinear optical response.  By comparison, for the standard model of avalanche ionization for which $\Delta\omega(t)\propto I(t)$, as seen by using Eq. (\ref{standard}) in (\ref{Del_om}), the peak of the blue-shift must lie within the temporal extent of the exciting pulse.  Here we see that the memory effects can greatly reduce the effectiveness of the plasma blue-shifting in the vicinity of the peak of the first pulse.  This is the case since the plasma takes time to build up from the initial low value.  In contrast, as shown in Fig. \ref{figure2} the lower intensity second pulse can experience a much larger plasma blue-shift in the vicinity of its peak: This is a consequence of the seed ionized density from the first more intense pulse which allows the plasma density and associated blue-shifting for the second pulse to build up much more rapidly.

The difference in plasma blue-shift experienced by the two pulses could also potentially serve as a diagnostic to measure the memory effects, and we remark that we obtained qualitatively the same blue-shift results in Fig. (\ref{figure2}) from the one-dimensional propagation simulations (not shown).  Here, however, we concentrate on the variation of the seed density with pulse delay as a diagnostic as it leads to a more direct experimental signatures in the propagation simulations.

\vspace{0.5cm}

\subsection{One-dimensional simulations}\label{IVC}

\begin{figure}[!h]
  \centerline{
    \includegraphics[clip,width=0.8 \columnwidth]{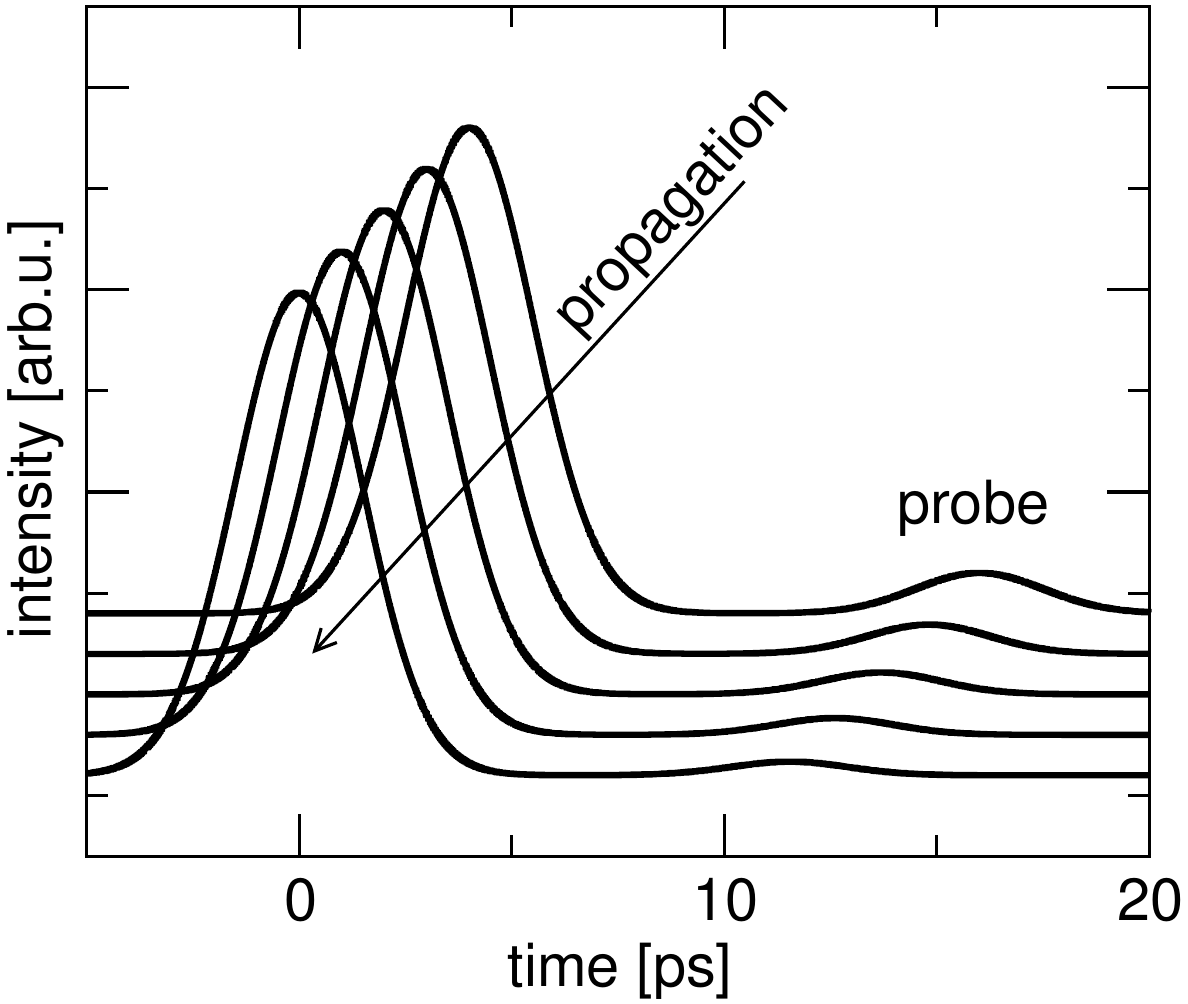}
  }
  \caption{\label{figure3}
   Raw data from the one-dimensional simulations showing the intensity profile of the propagating double-pulse for a variety of propagation distances: The arrow indicates increasing propagation distance, the pump pulse being on the left and the probe on the right.  
   }
\end{figure}

Building on the point model simulation shown in Fig. \ref{figure1}, we have performed one-dimensional simulations (no transverse variations) to show that the memory effects alluded to in Sec. \ref{IVB} remain qualitatively unchanged when propagation effects including plasma absorption and pulse distortion are accounted for.  These simulations are representative of what can be expected for propagation in a hollow core fiber filled with atmospheric air.  In particular, we have simulated the same double-pulse scenario as in Fig. \ref{figure1}, the first or leading pulse playing the role of a pump while the second weaker pulse acts as a probe that interacts with the electrons generated {\em inside and after} the pump pulse.  The raw data from these simulations is presented in Fig. \ref{figure3} which shows the intensity profile of the propagating double-pulse for a variety of propagation distances: The arrow indicates increasing propagation distance, the pump pulse being on the left and the probe on the right.  Since this simulation neglects transverse variations it is safe to say that observed losses are solely due to the electron generation. Then contrasting the pump and probe pulse intensity profiles indicates that the probe interacts with and/or generates more free electrons than the pump, as implied by the fact that the probe intensity suffers a larger relative decrease in intensity in comparison to the pump.  This is possible since the pump pulse initiates plasma generation that continues after it terminates, whereas the probe pulse experiences this electron population as a seed for further avalanche.  Thus, the probe can experience larger losses than the pump as indicated in Fig. \ref{figure1}.

\begin{figure}[!h]
  \centerline{
    \includegraphics[clip,width=0.8 \columnwidth]{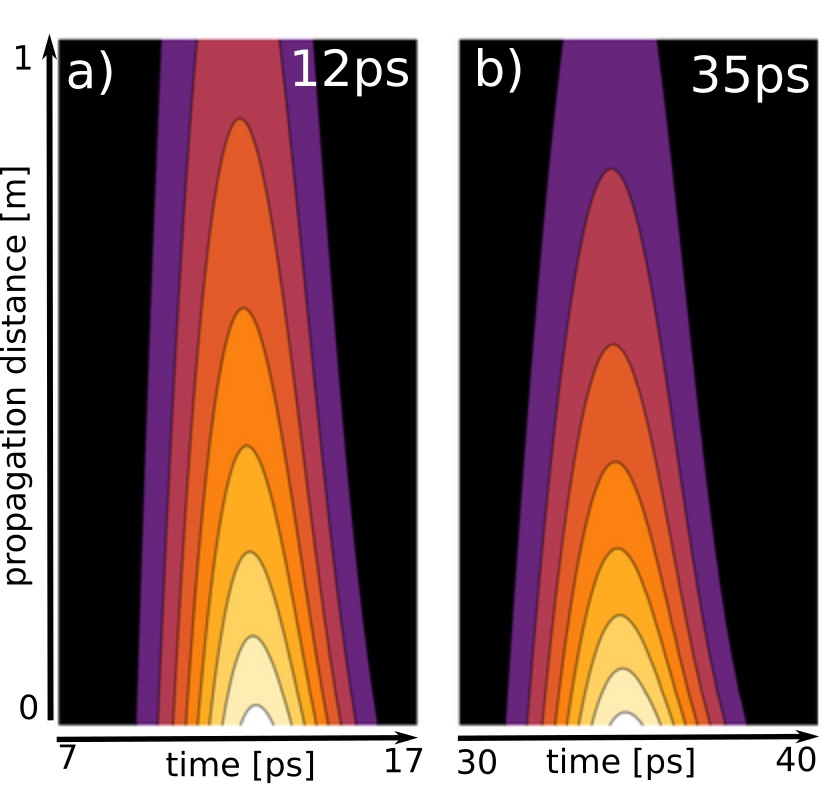}
  }
  \caption{\label{figure4}
    (Color online)  Color-coded intensity profiles $I(z,t)$ of the propagating pulses, with $z$ the propagation coordinate and $t$ the reduced time moving at the group velocity, for pump probe delays of a) $12$ ps, and b) $35$ ps.  }
\end{figure}

A key feature of our model is that the electron density can continue to grow even after the pump pulse has terminated.  This implies that if one increases the delay between the two pulses, the probe must experience a larger plasma density and hence larger absorption. This effect is illustrated in Fig.~\ref{figure4} for delays of a) $12$ ps, and b) $35$ ps: Each plot shows the color-coded intensity profile $I(z,t)$ of the propagating pulse, with $z$ the propagation coordinate and $t$ the reduced time moving at the group velocity.  What is revealed is that for the larger delay in Fig.~\ref{figure4}(b) the peak intensity decreases more rapidly with propagation direction implying a larger generated electron density.  One can also see that the trailing edge (right) of the probe pulse experiences greater losses than the leading edge, a sign even a weak probe can accelerate the avalanche generation of electrons.
    
\begin{figure}[!h]
  \centerline{
    \includegraphics[clip,width=0.8 \columnwidth]{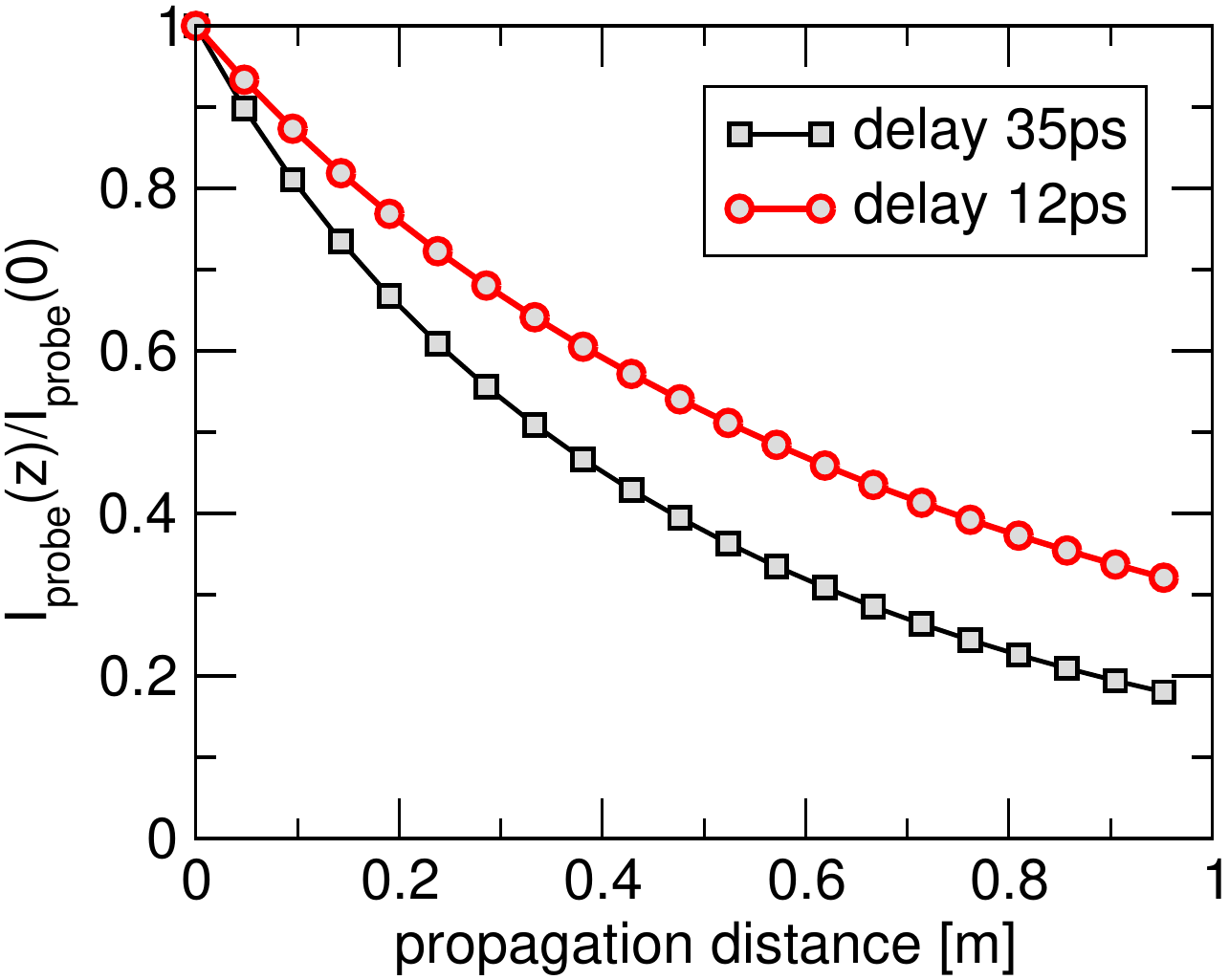}
  }
  \caption{\label{figure5}
    (Color online)  Peak probe intensities, normalized to the peak input value $I_p=0.12\times 10^{12}$ W/cm$^2$, of the propagating pulses versus propagation distance for the two different delays of $12$ ps, and $35$ ps.  
  }
\end{figure}

In order to quantify the difference in probe absorption, Fig.~\ref{figure5} shows the peak probe intensities, normalized to the peak input value $I_{probe}(0)=0.12\times 10^{12}$ W/cm$^2$, of the propagating pulses versus propagation distance for the two different delays.  Here we see that the longer delay leads to larger probe absorption, whereas according to the standard model in which electron generation does not occur between the pulses, this dependence on delay is not possible.


\subsection{Three-dimensional simulations}\label{IVD}

Next, we want to illustrate that the memory effects survive in a realistic situation taking full account of the spatio-temporal dynamics of the pump and probe pulses. The example that follows can be viewed as a proposal for a conceptually simple experiment designed to detect the memory effects in plasma generation by long-wavelength infrared pulses.

For this purpose, we consider the following variation on the double-pulse setup: The pump is once again the initiator of the plasma and we take it to be tightly focused in the transverse plane, whereas the probe beam is collimated. In particular the probe beam is taken to have a transverse width much larger than that of the plasma filament created by the pump pulse, both beams being coaxial:  Here we use a pump beam focused (linearly) to a width of $1.2$~mm, leading to a plasma filament width of $\sim 0.8$~mm, and a collimated probe width of $19$~mm.  Then under propagation the probe beam will experience plasma absorption at its center, leading to depletion of the central region of the probe and subsequent diffraction of this narrow hole (compared to the probe width).  As discussed in Ref. \cite{pavel}, this results in a transverse ring pattern, resulting from diffraction of the central hole, that can be detected in the transverse fluence profile. Here we assume that the probe is the second harmonic of the leading pulse, and that after the focus the fundamental is filtered out, while the fluence profile of the probe beam is recorded.
\begin{figure}[!h]
  \vspace{3mm}
  \centerline{
    \includegraphics[clip,width=0.95 \columnwidth]{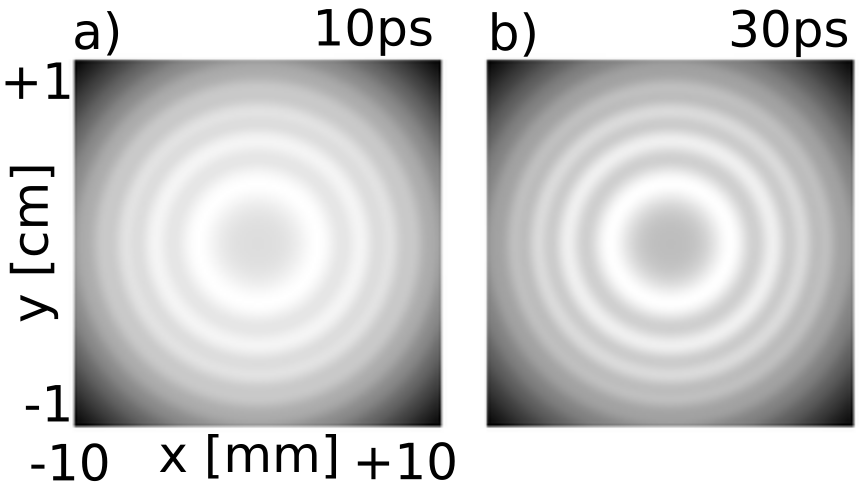}
  }
  \caption{\label{figure6}
   Simulated fluence profiles over the transverse plane $(x,y)$ for two different pump-probe delays of a) $10$ ps, and b) $30$ ps, and a propagation distance of one meter.   }
\end{figure}

Figure~\ref{figure6} shows simulated fluence profiles over the transverse plane $(x,y)$ for two different pump-probe delays of a) $10$ ps, and b) $30$ ps, and a propagation distance of one meter. The fluence maps are recorded over the 2$\times$2 centimeter central region of the beam and demonstrate that the longer $30$ ps delay leads to a more pronounced interference pattern as a result of the larger plasma density.  Once again, this is a manifestation of the memory effects predicted by our model.

\begin{figure}[!h]
  \centerline{
    \includegraphics[clip,width=0.8 \columnwidth]{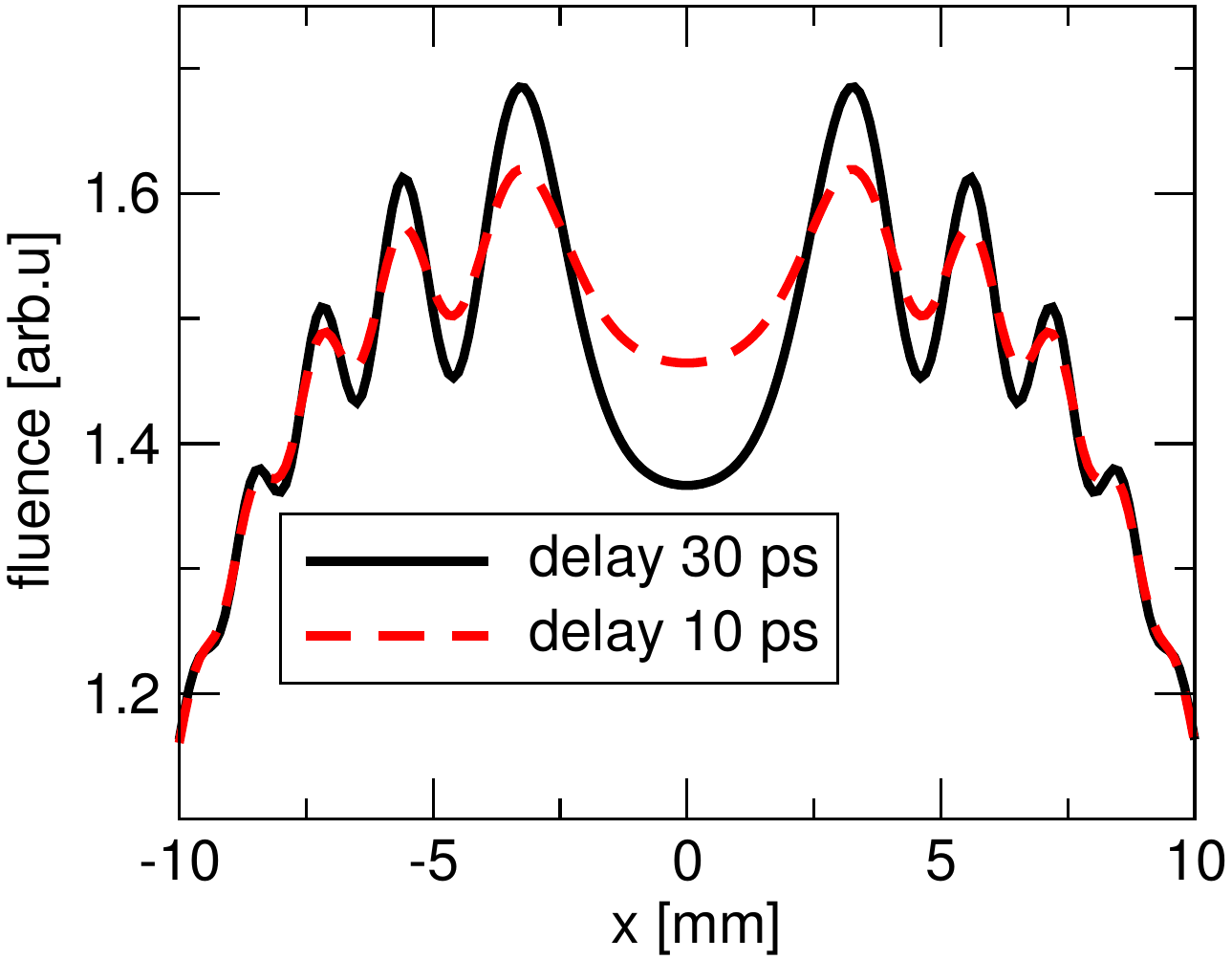}
  }
  \caption{\label{figure7}
    (Color online)  Lineouts of the fluence spatial profiles from Fig.~\ref{figure6} .  The solid black line is for a pump-probe delay of $30$ ps and the red dashed line is for $10$ ps.}
\end{figure}

For a more quantitative view, Fig.~\ref{figure7} shows lineouts of the fluence spatial profiles from Fig.~\ref{figure6}. The contrast seen in the fluence plots is about ten to twenty percent of the total, which should be easily detected in a real experiment.

We remark that an even simpler version of this experiment could result if both pump and probe have the same center wavelength:  Because the leading pulse diffracts quickly after its focus, the detected is mainly due to the probe beam. It is therefore not strictly necessary to probe at a wavelength different from the pump.

\section{Recurring issues}\label{issues}
After presenting our hot electron model of avalanche ionization for long-wave infrared picosecond pulses it behooves us to highlight where key issues still arise and summarize our responses.  In particular, these are issues that spark recurring discussions at conferences and elsewhere.
\begin{itemize}
\item[(1)]
{\bf Temperature relaxation time:}  This is by far the most controversial issue that arises.  Based on previous simulations of the quantum Boltzmann equation we have identified the time scale on which the nonequilibrium temperature $T_{kin}(t)$ approaches the equilibrium plasma temperature $T_{pl}(t)$.  That this timescale is given by the inverse plasma frequency is a well documented and supported conclusion, but the flawed intuition persists that Coulomb and quantum effects in weakly ionized atmospheric gases should be negligible.  The basic argument is that for such low electron densities there is no possibility for the collective effects that would allow for the introduction of a plasma frequency, but this misses the key point that the Coulomb potential is basically unscreened by virtue of the low density.  This issue can only be resolved using the quantum Boltzmann equation simulations that we have already performed.
\item[(2)]
{\bf Two temperature theory:}  It is often asked if the hot electron model is simply another two-temperature model but this is not the case.  In the hot electron model the time-dependent state of the system is described by the density of freed electrons $N(t)$ and the non-equilibrium temperature $T_{kin}(t)$ that is a measure of the mean kinetic energy of the electrons.  On the other hand, the plasma temperature $T_{pl}(t)$ is the equilibrium temperature towards which the system will relax, there is no implied sub-population of electrons at this temperature.  Thus, in distinction from the usual two-temperature type model where two population components are identified, each with its own temperature and density, our model is a one-component model with the plasma temperature acting as a reference temperature to which the system must evolve due to the varieties of Coulomb scattering.
\item[(3)]
{\bf Avalanche versus cascade ionization:}  It is known that the electron density can continue to rise even after the exciting pulse has passed.  For this reason the terminology has evolved that avalanche ionization is the process of electron multiplication during the pulse, whereas after the pulse it is referred to as cascade ionization.  From the perspective of our hot electron model no such distinction is needed by virtue of the memory effect that allows for electron multiplication after the pulse has passed, and we use the term avalanche ionization to cover both.  In particular, cascade ionization is then a manifestation of the memory effects.
\item[(4)]
{\bf Inclusion of excited states:}  It is often pointed out that our hot electron model allows only for impact ionization processes.  This is correct and it is well known that impact excitation of excited atomic states can also occur and reduce the amount of ionization that occurs.  To address this issue will require an extension of the current model to go beyond the freed electron density and allow for the excited state populations \cite{GaoPatSch17}.
\item[(5)]
{\bf EID model versus tunneling ionization:}  For our simulations for picosecond $10~\mu$m pulses we have used our model based on excitation-induced dephasing (EID) to calculate the laser-induced absorption as opposed to the more familiar tunneling-ionization mode.  Indeed measurements of tunneling ionization have been reported from the $10.6~\mu$m to the visible \cite{Ilkov,Walsh,LaiXuSza17} so why use the EID model?  The reason is that the previous measurements and theory were for ultra-low pressures, $10^{-7}$ Torr or less, whereas we consider atmospheric gases.  In this case the laser-induced ionization due to EID exceeds the tunneling ionization by orders of magnitude \cite{SchKolWri17}. 
\end{itemize}

\section{Summary and conclusions}\label{Summary}
In this Report of Progress we presented a hot electron model described by Eqs. (\ref{A})-(\ref{C}) that provides a microscopically motivated foundation for avalanche ionization in gases.  We have developed this model in response to the need for a reliable yet simple model that can capture the features of avalanche ionization for long-wavelength infrared pulses of picosecond duration that are beyond the standard model.  The key point is that avalanche ionization is driven by the temperature difference between the hot electrons and the ambient plasma, which in turn is driven by the light-field intensity.  As a consequence of this sequential coupling, the electron system retains a memory of the laser excitation.  An example of these memory effects and associated plasma blue-shifting is presented for picosecond LWIR pulses in air, where temperatures of the order of $10^6$ K were found.  For corresponding simulations of NIR pulses much lower temperatures are obtained, $\sim 10^4$ K, and as a result the memory effects are much less pronounced signaling that the effects described here are very important for LWIR but will have a much smaller effect for optical filamentation of NIR pulses.  

Propagation simulations in one- and three-dimensions are presented for double-pulses to both illustrate how the memory effects may be measured experimentally, and also demonstrate that the hot-electron model is readily incorporated into long distance pulse propagation codes.  The goal of these simulations is to point to diagnostic experiments that could be used to test the ideas put forward here, in particular to probe differences with the standard model.  For these simulations, the key memory signature is that the electron density continues to rise between the in the field free region between the pump and probe, in contrast to the standard model where this cannot occur. We anticipate that as the developments of this paper unfold and are further explored both theoretically and experimentally, they will have a direct impact on current efforts to produce optical filamentation in the LWIR region and beyond.

\section*{Acknowledgements}
This material is based upon work supported by the Air Force Office of Scientifc Research under Grant No. FA9550-16-1-0088, and partial support under an ONR MURI grant \# N00014-17-1-2705.  The Marburg work is funded by the Deutsche Forschunggemeinschaft (DFG) via SFB 1083.  EMW thanks SFB 1083 for support during his visit to the University of Marburg.  The authors thank Prof. Sergei Tochitsky for helpful comments, and Prof. Dmitri Romanov for discussions regarding impact ionization cooling.

\end{document}